%% file: Stereotype.tex
\documentclass[sigconf]{acmart}
\usepackage{booktabs} 
\usepackage{balance}
\DeclareMathOperator{\E}{\mathbb{E}}
\usepackage{mathtools}
\usepackage{graphicx}
\usepackage{subfig}
\usepackage{lscape}
\usepackage{float}
\usepackage{enumitem}

\setcopyright{none}
\acmConference{KaRS 2019 Second Workshop on Knowledge-Aware and Conversational Recommender Systems} {November 3rd-7th, 2019}{Beijing, China.}
\acmDOI{}
\acmISBN{Copyright for the individual papers remains with the authors. Copying permitted for private and academic purposes. This volume is published and copyrighted by its editors.}
\copyrightyear{2019}
\acmPrice{}

\pdfoutput=1

\begin{document}

	\title{Generating Stereotypes Automatically For Complex Categorical Features}

	\author{Nourah A.~ ALRossais}
\affiliation{\institution{King Saud University, KSA}{University of York, UK}}
\email{nar537@york.ac.uk}
	
\author{Daniel Kudenko}
	\affiliation{\institution{University of Hannover, Germany}}
	\email{kudenko@l3s.de}
	
		\renewcommand{\shortauthors}{N. ALRossais et. al.}

	\begin{abstract}
	
	 In the context of stereotypes creation for recommender systems, we found that certain types of categorical variables pose particular challenges if simple clustering procedures were employed with the objective to create stereotypes. 
	A categorical variable is defined to be complex when it cannot be easily translated into a numerical variable, when the semantic of the categories potentially plays an important role in the optimal determination of stereotypes, and when it is also multi-choice (e.g., each item can be labelled with one or more categories that may be applicable, in a non pre-defined number).
	The main objective of this paper is to analyse the possibility of obtaining a viable recommendation system that operates on stereotypes generated directly via the feature's metadata similarities, without using ratings information at the time the generation of the classes. The encouraging results using integrated MovieLens and Imdb data set show that the proposed algorithm performs better than other categorical clustering algorithms like k-modes when clustering complex categorical features. Notably, the representation of complex categorical features can help to alleviate cold-start issues in recommender systems.




	\end{abstract}


	\maketitle
	
	\input{introduction}
	\input{relwork}

	\input{Correlation}

\input{hierarchical}

	\input{algorithm}

\input{Comparison}

\input{rating}

	\input{conclusion}

	\bibliographystyle{ACM-Reference-Format}
	\balance
	\bibliography{Stereotype}
	
\end{document}

%% file: introduction.tex
\section{Introduction}
\label{sec:introduction}
The growing importance of recommender systems has motivated the research community to look for diverse techniques and approaches to solve challenges like the "new user" or "new item" problems. A promising approach for improving recommendations in the cold start phases for the new user/item problem is stereotype based modeling. Rich \citep{rich1979user} was the first to propose the utilization of stereotypes in user modeling and recommender systems. A stereotype depicts a collection of attributes that are relevant for a collection of users (or items) \citep{lock2005performance}.

Historically, in the pioneering works on stereotyping, the majority of the stereotype classes were built manually by the operator with knowledge of the problem and data, see for example \citep{rich1979user,kay1997learner,kay1994toolkit,brajnik1994shell}. This approach has obvious limitations like the operator building classes manually may miss or disregard important relationships (features) that effectively classify and define a stereotype. It is therefore paramount to create a stereotype building procedure that assembles the classes in a systematic manner. 

A machine learning based study should address the design of the optimal data representation to tackle the problem at hand, and the same applies to stereotyping; a deeper understanding of the underlying features that are involved in the generation of the stereotypes will help design better algorithms. The stereotyping process is made more challenging, but at the same time more representative of real-world applications, by the presence of categorical variables which are multi-entry in a non-strict sense, i.e. the number of multiple labels describing a category for an item/user is not pre-specified. In some ways these variables can be viewed as multiple choice answers to a questionnaire, with the underlying idea of "pick all that applies". For example in the MovieLens and Imdb data sets, ~\citep{harper2016movielens}, a typical complex categorical feature is constituted by the movie genre: an item may be just "drama", another may be "drama" plus "romance" plus "historic". 

 Complex categorical features - especially when treated via a naive one to enne encoding, and when their dimensionality is large - ends up constituting the features that dominate the stereotype creation via clustering techniques; this implies that these types of features should be potentially addressed in a different manner in the context of stereotyping. The present paper main contribution is to demonstrate how item stereotypes can be built automatically for complex categorical features in a way that is independent of user's rating. Ultimately, demonstrating how the use of stereotypes can be effective during the cold start phase.


The rest of the paper is organized as follows: Section ~\ref{sec:relwork} summarizes related work. A correlation matrix analysis leading to discover stereotypes automatically is presented in Section~\ref{sec:Correlation}. An experiment to benchmark our algorithm with k-modes as well as application of stereotypes to recommendation system is carried out in Section~\ref{sec:comparison} and ~\ref{sec:rating}. Finally, conclusion and future work are given in Section~\ref{sec:conclusion}.


%% file: relwork.tex
\section{Related Work}
\label{sec:relwork}

Clustering based algorithms applied to a data set describing items can provide either a direct representation of stereotypes or provide valuable insights in what features are most distinctively driving class separations. The main challenge in the application of a clustering algorithm to the general problem of extracting classes from the data resides in the standardization of the data. In the most common scenario mixed numerical and categorical features are present; in addition, extra complexity may arise by categorical features that are not simply labels but may require machine-based insights in the language in which they are expressed.

Standard clustering algorithms, like the well-known k-means and its variations, discover structures in the data by working with Euclidean distances, and minimizing the total variance of the distances between the cluster's centroids and the individual data points, see \citep{gan2007data}. For categorical features the concept of distance, and of ordering in general, may be difficult to define and, when not meaningless, it may introduce unexpected false relationships.

A body of research exists for the application of clustering concepts to categorical data, in \citep{huang1998extensions} the k-mode algorithm was introduced to deal with categorical data. In the k-modes algorithm the centroid of clusters is no longer identified by means but with modes, a matching dissimilarity measure is introduced to deal with categorical objects, and the clustering cost function is minimised using a frequency-based method to update the modes. Several marginal improvements have been introduced to k-modes, see for example \citep{sangam2015k,cao2012dissimilarity}, where the improvements are all directed toward the formation of the dissimilarity measure used in k-modes. In \citep{he2006approximation} similarity and efficiency of k-mode is investigated and related to the k-median approach. 

The authors in \citep{cao2017algorithm} suggested an algorithm for clustering categorical data with set-valued (i.e. complex value). However, the algorithm is relatively complex and fails to consider the effect of correlation between labels embedded in the data.

It is paramount to create a mechanism to handle complex categorical features. To the best of the authors' knowledge all possible categorical clustering algorithms available (like k-modes and its possible variations) have been developed for single choice categorical variables, and hence they cannot be applied directly when there are multi-choice categorical features. Additionally, the majority of the clustering, similarity metrics, as well as dimensionality reduction approaches operate on the users to items rating matrix. The present work analyses a different aspect, namely the possibility of obtaining a viable recommendation system that operate on stereotypes that are generated directly via the feature's metadata similarity, without using the rating at the time of the generation of the classes.


%% file: Correlation.tex
\section{Constructing Stereotypes For Complex Categorical Features}
\label{sec:Correlation}

For a typical multi-entry categorical feature that describes a given item, there will be a number of entries where multiple labels are assigned to the same item. By investigating a large enough set of items with multiple entries one can extract what type of relationships exist, if any, between pairwise labels. This can be done by investigating the correlation matrix of the encoded multi-entry feature. 
The use of correlation, and in particular of the correlation matrix between pairwise labels, which is defined starting from the covariance matrix, is supported by the objective to search for intrinsic relationships between the labels that are present in the sample data. For a revision of the basic statistical concepts behind covariance and correlation see for example \citep{tsokos}.

The first step consists of converting the categorical feature in a \textit{multi}-one-hot encoding, and then computing the correlation matrix between categories. The word \textit{multi} in front refers to the fact that each item can be specified with one or more (pick all that apply) categories so the encoding has as many ones. The correlation matrix can be defined in a standard way; given a multi-one hot encoded observation for the multi-entry categorical feature, $x_i$ for $i=1, \dots, N$ possible categories, the covariance matrix is defined as:

\begin{equation}
CV_{i,j}= \E{[(X_i - \mu_i) (X_j - \mu_j)]}  \qquad \quad i,j=1,	\dots, N
\end{equation}

\begin{equation}
\mu_k = \E{[X_k]}  \nonumber
\end{equation}

Where $\E[\quad]$ is the expected value operator. From the covariance matrix, the correlation matrix $R$ is obtained using the product of the standard deviations along directions $i,j$ as a normalization coefficient:

\begin{equation}
R_{i,j}= \frac {CV_{i,j}} {\sigma_i\sigma_j} 
\label{EQU:correlation_matrix}
\end{equation}

\begin{equation}
\sigma_k^2= \E{[(X_k - \mu_k)^2]} \nonumber   
\end{equation}

The values in the correlation matrix~\ref{EQU:correlation_matrix} would already suggest, for the cases where the labels are not many, which categories can better be coupled with each other. To enhance further the grouping between categories, one can group (also improperly called clustering) entries of the correlation matrix that are most related between each other. Several algorithms have been proposed in the literature, see for example \citep{friendly2002corrgrams} and \citep{Zimek2008correlation} references therein, most of these revolve around the application of hierarchical clustering using the correlation matrix entries to define a penalty distance function. The penalty function can be introduced in several different ways, in this context the following definition is adopted

\begin{equation}
P_{i,j}= 1- |R_{i,j}| 
\label{EQU:Penalty}
\end{equation}

Which constitutes a simple linear penalty: low correlations around 0 are penalized more than high positive or negative correlations (near +/- 1). 

For this research a greedy grid search algorithm was developed to rank possible permutations of columns and rows in the correlation matrix to gain an initial understanding of the grouping in the two large and significant complex features: "movie genre" and "keywords" in the MovieLens/Imdb data sets \citep{harper2016movielens}.

\begin{figure}
  \centering
  \includegraphics[scale=.33]{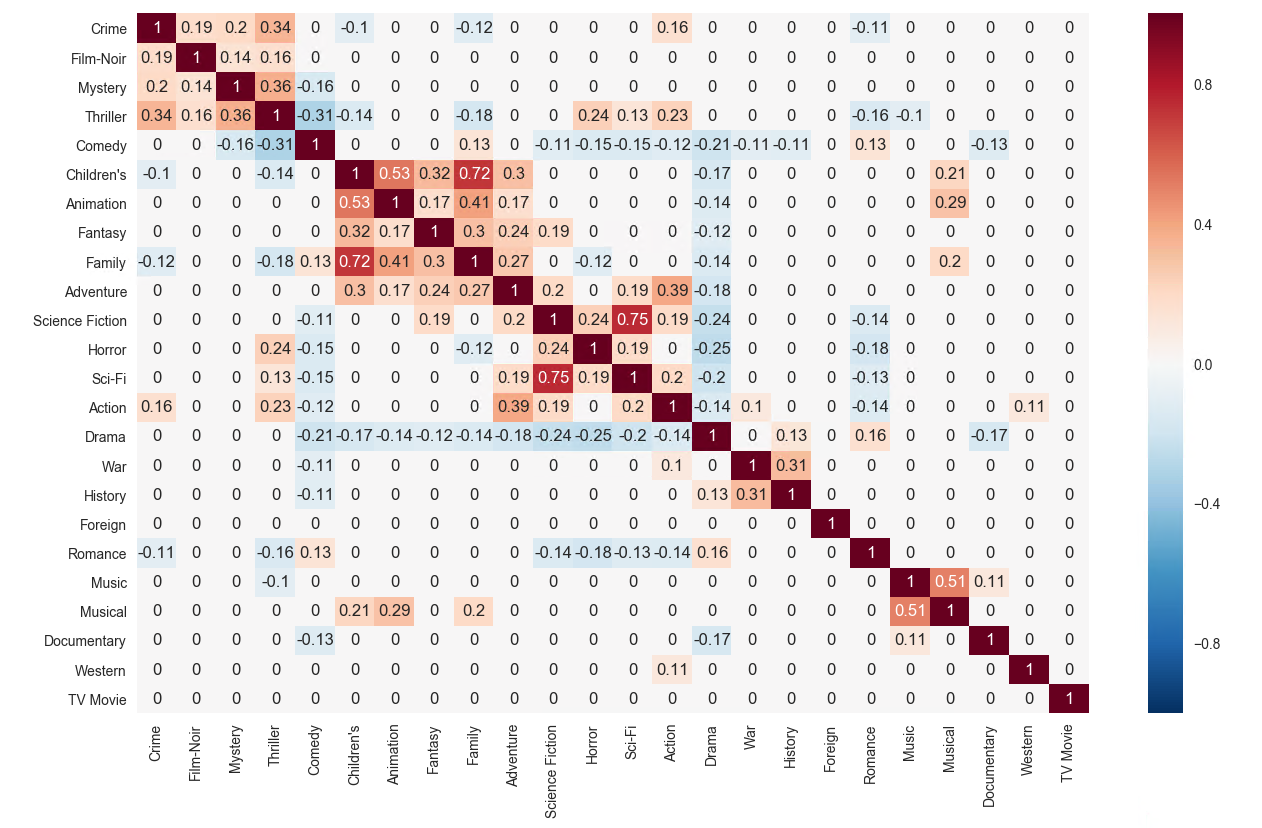}\\
  \caption{\small{Correlation matrix for the genre feature} }\label{FIG:Correlation_genre_clustered}
  \vspace*{-3mm}
\end{figure}

When the greedy grid search is applied to the row/columns permutations of the "genre" feature, the correlation matrix shown in Figure~\ref{FIG:Correlation_genre_clustered} is obtained. The permutations make it easier to identify "groups" that can be considered as clusters for that feature, groups which we will refer to as stereotypes. In the case at hand for the feature "genre" we can see that the first group is constituted by ("Film-Noir", "Thriller", "Crime" and "Mystery"). A second group is constituted by ("Children's", "Animation", "Family", "Fantasy") etc. 

A similar correlation analysis was performed for the feature "keywords" by restricting the attention to keywords that appear in the training data a sufficient number of times, in order to avoid fitting any pattern to keywords that are rarely used. In this context, only keywords that were used at least   20 times over the items of the data were retained. 
Keyword correlations are in general much weaker than those for genre, as a result of the vast "scattering" over hundreds of keywords over the data. 

%% file: hierarchical.tex
\subsection{Hierarchical
Clustering}
\label{SEC:hierarchical}
The simplified analysis conducted via the permutation algorithm motivates the use of a more rigorous clustering approach for the definition of stereotypes in complex categorical features. 
Following \citep{Zimek2008correlation}, in order to apply formal clustering to a correlation matrix, it is necessary to introduce both a "metric" that defines distances between pair of observations, and a "linkage" criterion whose role is to define the similarity/dissimilarity across groups (clusters) of observations based on the distances between single observations. The metric/distance needs to respect the properties: \newline
a) positive defined, b) elements that are nearer to each other have lower distance than elements that are further apart from each other.

To be able to respect these properties starting from the definition of correlation, the latter needs to be inverted, in the sense that the closer the correlation to $|1|$, the smaller the distance, with the limiting case of correlation going toward +1 (-1) and distance approaching 0. Such an inverted correlation metric can be obtained in several different ways, and it is often called dissimilarity measure, see \citep{podani2000introduction} for examples. Two such ways to introduce the dissimilarity from the correlation are the linear and the quadratic:



\begin{equation}
    D_{i,j}^{(a)}=1-|R_{i,j}|
    \label{EQ:diss1}
\end{equation}

\begin{equation}
    D_{i,j}^{(a)}=\sqrt{1-R_{i,j}^2}
    \label{EQ:diss2}
\end{equation}

As discussed, the dissimilarity measures need to be complimented with a linkage criterion. In the hierarchical clustering literature there are many alternative linkages proposed, see \citep{friedman2001elements} for a general review. In this research the most widespread and general linkages will be employed, the \textbf{single}, \textbf{complete} and \textbf{Ward} linkages. 

The two metrics for dissimilarity~\ref{EQ:diss1} and ~\ref{EQ:diss2} are applied in a hierarchical clustering algorithm via the three alternative linkages criteria to the correlation matrix for the feature "genre" and "keywords". Investigation of the results leads to several general observations. The first observation is that, regardless of the linkage adopted, dissimilarity metric~\ref{EQ:diss2} tends to compress excessively toward 1.0 entries that have low correlations (below 0.4 - 0.5 in absolute value). The spectrum of correlation values between 0 and 0.5 will lead to a dissimilarities between 0.8660 and 1.0. When these dissimilarities are investigated via the dendrogram formation the resulting dendrograms appear to be too compressed for matrices that tends to have average low correlations in magnitude. Measure~\ref{EQ:diss2} therefore is more suited for exploring situations where the correlations tend to be high in average over the correlation matrix, above 0.4 - 0.5 in absolute value. As a rule of thumb there should not be more than a handful of pairwise correlation entries above 0.4 to use measure ~\ref{EQ:diss1}. For this reason, attention in this paper will be focused to just dissimilarity~\ref{EQ:diss1} for all linkages.

The logical grouping identified by single linkage and dissimilarity~\ref{EQ:diss1} (not shown) - seems to be less congruent with a "human" subjective assignment. A metric could be used to evaluate the hierarchical links discovered, see for example the Fowkles and Mallows metric \citep{fowlkes1983method}, however such a metric requires the availability of the true labels for the clusters. In addition to the fact that the true labels are not available, even if they were available for complex categorical features like genre and keywords, the true labels would reflect the "expert opinion" of the operators defining the labels - hence they would also be subjective rather than objective. For this very reason evaluation of the clustered correlation matrices and dendrograms is performed via a subjective judgment at this stage of the research.

\begin{figure}
  \centering
  \includegraphics[scale=.30]{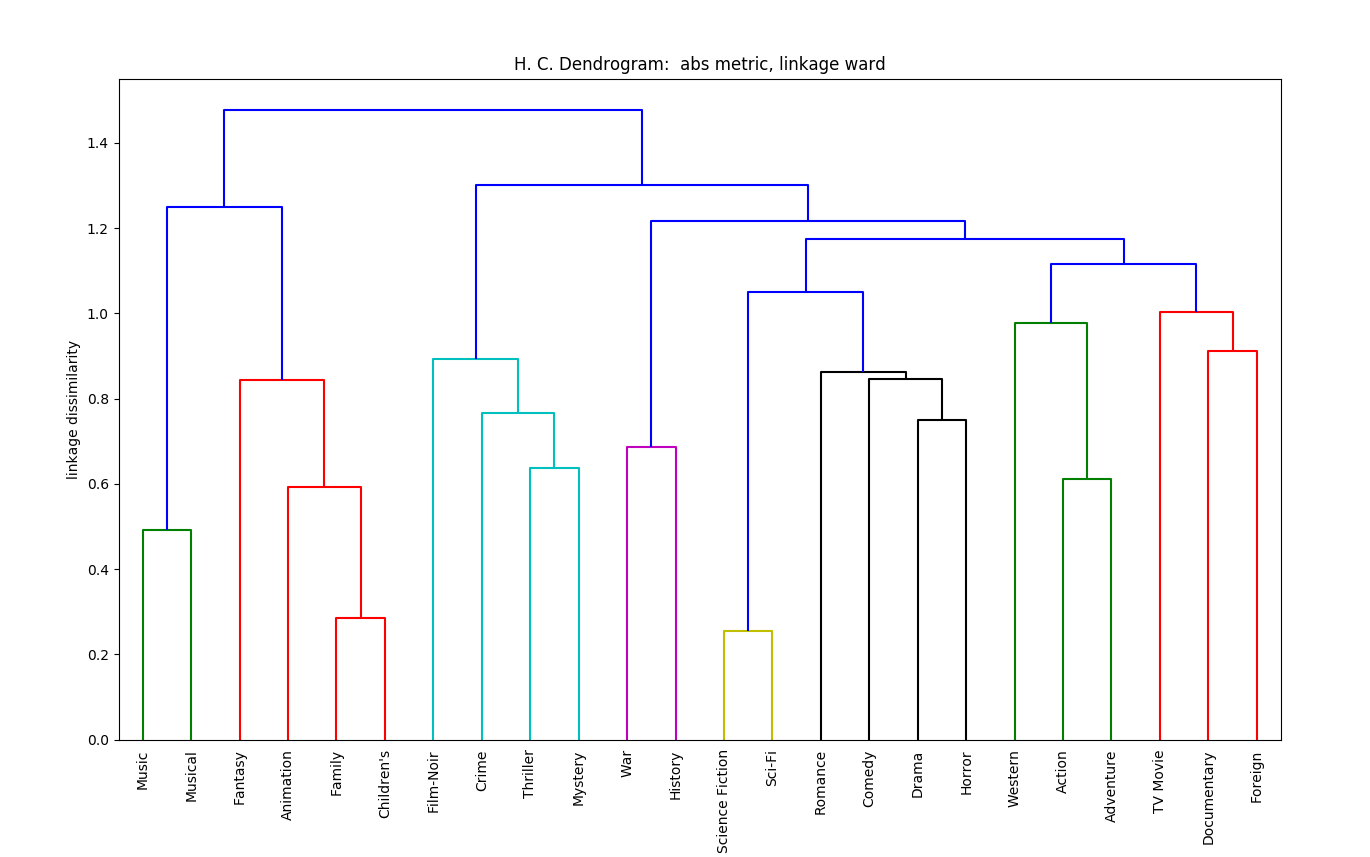}
  \caption{\small{Genre dendrogram using metric~\ref{EQ:diss1} and Ward-linkage}}%
    \label{FIG:diss1_ward}
    \vspace*{-5mm}
\end{figure}

The groupings provided by the complete linkage (no shown) and Ward linkage as shown in Figure~\ref{FIG:diss1_ward}, are overall similar. However, the Ward linkage seems to better represent the hierarchical links between different levels of the groupings. 

The same hierarchical clustering of the correlation matrix, and dendrogram study of the linkage dissimilarity splits, is then applied to the other complex categorical feature of the sample data under investigation: "keywords". It was previously observed, Section~\ref{sec:Correlation}, that the average in sample correlation between keywords falls in a similar range of values, if not lower, as the average correlation among categories of the genre feature. For such a reason given the previous observation about the fact that the dissimilarity measure~\ref{EQ:diss2} tends to compress toward 1.0 low absolute values of correlation and to provide a greater zoom on higher correlation values, measure~\ref{EQ:diss2} will be put aside in favour of dissimilarity measure ~\ref{EQ:diss1}.

In a similar fashion as that observed before, the complete linkage, and the Ward linkage, have proved to be better distinctions and grouping criteria. Confirming the findings from the analysis on the feature genre, it is possible to see how the two linkages form very similar groups across keywords dendrograms (not displayed), and how such groups can be "approved" from the subjective point of view of the categorization. 
However, the Ward linkage appears once more to be superior in defining a better vertical separation of the logical links in the dendrogram. This characteristic is extremely important when considering possible different levels for cutting automatically a dendrogram as a way to create logical groups for the labels of a complex categorical feature.

\setlength{\intextsep}{15pt minus 2pt}

%% file: algorithm.tex
\begin{figure}[H]
  \centering
  \includegraphics[scale=.4]{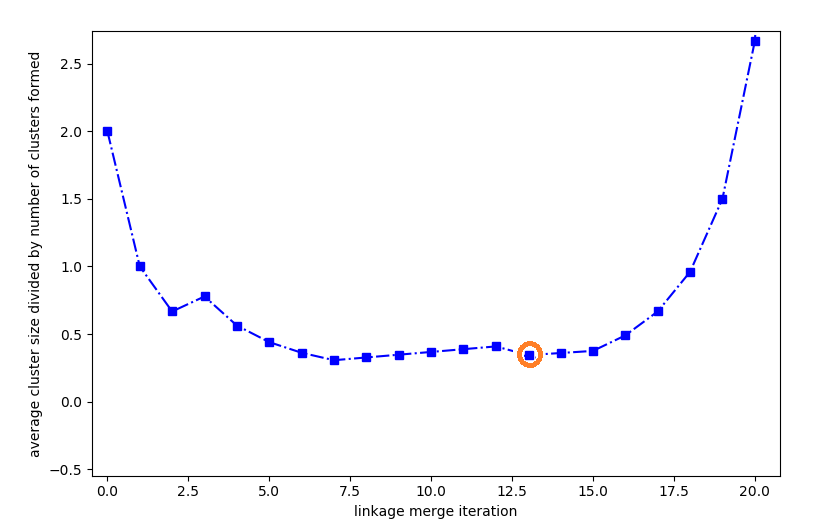}\\
  \caption{\small{The red circle in Dendrogram iteration ratio indicates the local minimum which is most to the right.}}\label{FIG:Genre_ratio}
  \vspace*{-5mm}
\end{figure}

\subsection{An Automatic Procedure for the Creation of Stereotypes}
\label{sec:algorithm}

It was observed in Section~\ref{SEC:hierarchical} that the vertical separation between the splits in the branches of a dendrogram, especially the one provided by Ward's linkage criterion, would be ideal for calibrating the height of a truncation in the dendrogram, thus obtaining as a result the groups of labels, hence the stereotypes. In order to accomplish the automatic stereotype creation, a systematic cutoff should be selected to decide at which height of the dissimilarity linkage one should truncate the dendrogram.

A dendrogram truncation criteria can be implemented by examining how the linkage merge iterations are shaping the clusters discovered from the bottom up (i.e. from the stronger links toward weaker links). As the iterations progress the number of clusters formed grows, then from a critical iteration onward, the structures discovered begin to merge toward a single cluster. This dynamic can be summarized by monitoring the average cluster size and the number of clusters formed up to a given iteration. The cut off procedure can therefore be implemented via a dual criterion: 
\begin{itemize}[noitemsep,topsep=10pt]
    \item By looking for the last maximum, or last local plateau in the number of clusters as a function of the iteration. 
    \item By applying a reverse elbow procedure to the average cluster size.
\end{itemize}

\begin{figure}[H]
  \centering
  \includegraphics[scale=.45]{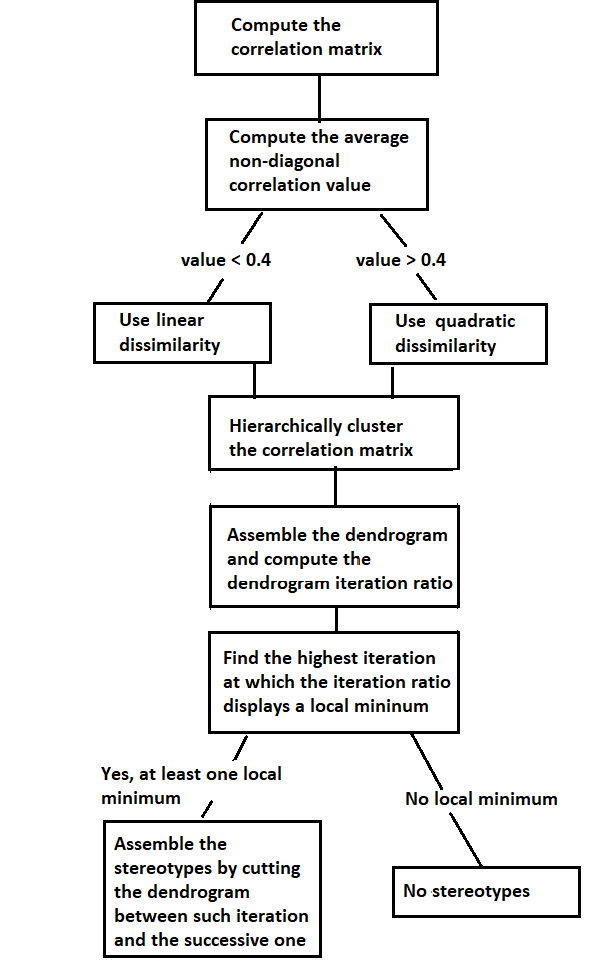}\\
  \caption{\small{Algorithm to assemble stereotypes for complex categorical features.}}\label{FIG:Sub_StereotypeAlgorithm}
  \vspace*{-5mm}
\end{figure}

The two criteria can also be coupled by taking the ratio, at any iteration, of the average cluster size divided by the number of clusters formed. This is shown for convenience in Figure~\ref{FIG:Genre_ratio} for the genre feature, and for simplicity such quantity will be referred to as \textit{dendrogram iteration ratio}. The cutoff procedure then reduces to finding the highest iteration exhibiting a local minimum in the iteration ratio. The only situation in which this idea would fail is in the case of a monotonically increasing dendrogram iteration ratio, that is found when there are no real underlying groups in the data, and the data is just grouped into an ever growing single cluster that will end up comprising the entire data set. In this special case the conclusion should be: the feature cannot be split into stereotypes. The algorithm to create stereotypes for complex categorical features is illustrated in Figure~\ref{FIG:Sub_StereotypeAlgorithm} for completeness. The application of the algorithm of Figure~\ref{FIG:Sub_StereotypeAlgorithm} to the genre and keyword features leads to the stereotypes of
Table~\ref{TAB:Genre_Sub_Stereotype}.

\vspace*{-3mm}

\begin{table*}
\small
  \centering
\begin{tabular}{|p{0.5 cm} p{6cm} p{7cm}|}
  
  \hline
  & \textbf{Stereotypes-Genre} & \textbf{Stereotypes-Keyword}   \\
  \hline 
  
  1 & ['Music','Musical']  & ['Violence','Explosion']     \\
  2 & ['Fantasy','Animation','Family','Children's'] & ['Nudity','Sex','Female nudity']\\ 
  3 & ['Action','Adventure',Western'] & ['Drug','Death',Murder','Police','Robbery'] \\
  4 & ['War','History']& ['Lawyer','Rape','Suspence','Serial killer','London England']\\
  5 & ['TV Movie', 'Documentary','Foreign'] & ['Film noir', 'Obession','New York']\\
  6 & ['Film Noir','Crime','Thriller','Mystery'] & ['Dying and death','Prostitute','Revenge']\\
  7 & ['Romance','Comedy','Drama','Horror'] & ['High School','Teenager','Suicide','Teacher']\\
  8 & ['Science Fiction','Sci-Fi'] & ['Independent film','Gay','Woman director']\\
  9 & & ['Based on novel','Biography','Kidnapping']\\
 10 & & ['Love','Friends','Jealousy','Adultery','Paris','Wedding']\\
 11 & & ['Sequel','Monster','Dystopia','Alien']\\
 12 & & ['Friendship','Father son relationship','Dog']\\
 13 & & ['Los angeles','Detective','Family']\\
 14 & & ['World war ii','Widow']\\
 15 & & ['Prison','Escape']\\
 16 & & ['Musical','Sport']\\

  \hline
  \end{tabular}
  \caption{\small{Stereotypes automatically generated using algorithm in Figure~\ref{FIG:Sub_StereotypeAlgorithm} for the feature: Genre and Keyword. }}\label{TAB:Genre_Sub_Stereotype}
    \vspace*{-5mm}
\end{table*}

%% file: Comparison.tex
\section{Comparisons of Our Algorithm With Other Clustering Algorithms}
\label{sec:comparison}

In this Section the clustering results (Table~\ref{TAB:Genre_Sub_Stereotype}) of the automatic stereotype algorithm suggested are compared with the results of the categorical clustering algorithm k-modes \citep{huang1998extensions}. K-modes is a clustering algorithm that is widely used in the literature. In order to create meaningful and useful stereotypes in the context of a recommender system, we are interested in an algorithm which is capable of grouping all the labels of the categorical feature under exam in stereotypes, without at priory excluding any labels.



In order to apply k-modes, for each possible label of the complex categorical feature under investigation, a new variable is introduced to represent a true/false encoding. This is not dissimilar from the concept of one to enne encoding, but in this context it is used to define not a numerical coordinate, but a multi-valued categorical representation instead.

The k-modes clustering algorithm can be initialized in different ways, for example following Huang \citep{huang1998extensions} the artefacts (e.g. the localization of the centroids) are placed in a random manner across the feature space, or following Cao \citep{cao2009new}, who suggested the artefacts to be placed in feature space based on initial density/frequency estimations. Once the method is initialized, the k-modes clustering implementation minimizes a cost function defined as the sum of all distances from each point to the cluster artefact that it believed to belong to. The concept of distance for categorical variables is defined via a delta function dissimilarity measure as described in \citep{huang1998extensions}.

\begin{figure}
  \centering
  \includegraphics[scale=.45]{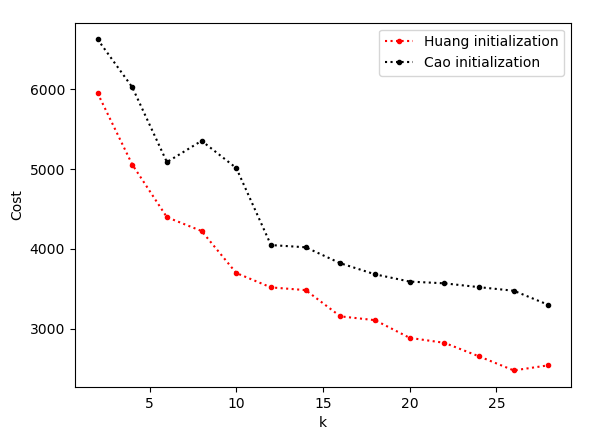}\\
  \caption{\small{Inverse elbow methodology applied to the k-modes clustering of the Genre feature.} }\label{FIG:Capture-score-genre-elbow}

\vspace*{-5mm}
\end{figure}

An inverse elbow methodology is applied to the cost function of k-modes as displayed in Figure~\ref{FIG:Capture-score-genre-elbow} for the genre feature partitioned via k-modes with both "Huang" and "Cao" initializations. Both cost functions decay with a lower rate of decay as the number of clusters, k, increases. However, it is not straightforward to identify a single well-defined kink in the decay graphs, and for this reason the results of the k-modes clustering for the genre feature are inspected by looking at the centroid characteristics at both (k=5) and (k=10). 

Table~\ref{TAB:5 centroides} displays the result for (k=5). The most interesting finding (also applied when k=10, not shown) is that the frequency approach underneath k-modes leads to absence of lower frequency labels. For example, genres like "War", "Western" and "Documentry" are not presented due to lower frequency in the item population. However, we argue that these labels should indeed be retained as they may represent specific niche users preferences, and are required in the recommendation items coordinates. Similar results were obtained concerning the application of k-mode for the feature keywords (not shown), and our empirical experience lead us to favor our algorithm over k-mode for the stereotype construction of complex categorical features. 
\vspace*{-3mm}


\begin{table*}
\small
  \centering

\begin{tabular}{|p{0.25 cm} p{6.75cm}|p{7.5cm}|}
  
  \hline

  \multicolumn{2}{|c|}{\textbf{Centroid Composition (Huang)}} & \textbf{Centroid Composition (Cao)}   \\
  \hline
  
 1&  ['Drama','Comedy','Romance'] & ['Family','Children's','Animation']     \\

 2 & ['Drama'] & ['Drama'] \\

3 & ['Adventure','Family','Children's','Animation']  & ['Comedy','Adventure','Family','Fantasy','Children's'] \\

4 & ['Comedy'] & ['Comedy']\\

5 &  ['Thriller','Action']
& ['Thriller','Action','Adventure','Science Fiction','Sci-Fi']\\
  \hline

\end{tabular}
  \caption{\small{K-modes resulting centroids composition for 5 clusters and the Genre feature.}}\label{TAB:5 centroides}
  
  \vspace*{-5mm}

\end{table*}

%% file: rating.tex
\begin{table*}[h]
\centering  \small
\caption{Baseline versus stereotype-based models}
\label{table:Summary}
    \begin{tabular}{|c|c|c|c|c|c|c|c|c|c|c|c|c|}
     \hline
    & \multicolumn{6}{|c|}{\textbf{New User}} &\multicolumn{6}{|c|}{\textbf{New Item}}
    
    \\\hline
   Metric
    & \multicolumn{2}{|c|}{\textbf{RMSE}} &\multicolumn{2}{|c|}{\textbf{MAE}}
    &\multicolumn{2}{|c|}{\textbf{Time in second}}
    
    &\multicolumn{2}{|c|}{\textbf{RMSE}} &\multicolumn{2}{|c|}{\textbf{MAE}}
    &\multicolumn{2}{|c|}{\textbf{Time in second}}
    
    \\\hline
     Method &  \textbf{Baseline} & \textbf{Stereo.}  &  \textbf{Baseline} & \textbf{Stereo.}
     &  \textbf{Baseline} & \textbf{Stereo.}
     &  \textbf{Baseline} & \textbf{Stereo.}&  \textbf{Baseline} & \textbf{Stereo.} &  \textbf{Baseline} & \textbf{Stereo.} \\ \hline
   \textbf{Linear Reg} & 0.940 & 0.939 & 0.743 & 0.743 & 10.702 & 8.545& 0.939 & 0.935 & 0.740 & 0.736 & 10.827 & 8.123 \\ \hline
   \textbf{NN Reg} & 0.918 & 0.906 & 0.724 & 0.712 & 69.5 & 53.4 & 0.928 & 0.918 & 0.735 & 0.728 & 56.87 & 45.11\\ \hline
   \textbf{XGBoostg} & 0.913 & 0.901 & 0.721 & 0.710& 90.57 & 50.75 & 0.926 & 0.918 & 0.738 & 0.729 & 90.568 & 49.192 \\ \hline
 \end{tabular}
   \vspace*{-2mm}
 \end{table*}

\section{Stereotype Based Recommendation Performance}
\label{sec:rating}

In the literature related to the application of recommender systems to predict cold start users' ratings on movies, the predictive algorithms applied most often aim to predict ratings. Our data set is assembled by combining the rating data from the MovieLens data set with the Imdb movie attributes. The combined data set contains 1,000,209 ratings, 6,040 users and 3,827 movies.
The rating is an integer value ranging from 1 to 5. Each movie can be categorised with 24 different genres and 71 different keywords (keywords that appear in the dataset more than 20 times). Generally, user-item ratings exhibit different kinds of global effects \citep{bell2007scalable}. For instance, some users always tend to give higher ratings on items than other users, as well as some items on average receive more positive user feedback than other items. In order to compute accurate rating predictions these bias effects (user and item biases) need to be removed from the data. Many techniques have been proposed in the literature, for example subtracting from the original entries user-average to remove individual user preferences and/or subtracting item-average to remove the item popularity effects \citep{spiegel2009hydra,bell2007scalable}.

In our study we normalize the ratings made by each user by converting them to standard scores:

\begin{equation}
  \tilde{r} = \frac{(r - \widehat{\mu_u})}
  {\widehat{\sigma_u}}
     \label{EQ:Normalize_Rating}
\end{equation}

Where $\widehat{\mu_u}$ is the mean rating per each user and $\sigma_u$ is the standard deviation per each user.

We generated two recommendation models: a "baseline model" which uses all features available in the data in the way they are provided (this is the model against which we benchmark our results), and a "stereotypes-based model" where complex categorical features have been stereotyped but all remaining features remain the way they are.
In order to measure the impact of replacing the original features with the stereotypes, and simulate cold start situations (new users/new items), the data set has been split into two alternative experimental sets:

\begin{itemize}
    \item Split A - For each item in the data set the models are trained on the preference set expressed by a subset of users (randomly selected). The remaining users are left out, and used to test the accuracy of the models; this method enables us to simulate the performance on "new users".
 \item Split B - For each user in the data set the models is trained on all preferences expressed for a subset of items (randomly selected). The remaining items are left out, and used to the accuracy of the models; this method enables us to simulate the performances on "new items".

\end{itemize}

For each of the two experiments we have tested several machine learning algorithms from the simplest method (Linear regression) to the most popular (Neural Networks and XGBoost) with the aim to improve model performances and to confirm that our conclusions related to the application of stereotypes do not depend on the model chosen. 

It should be noted that, for numerical regression style models, there is no guarantee that the rating will fall between 1 and 5, therefore the resulting prediction, when transformed back (rescaled) into a rating "r", must be capped/floored in the following manner: 

\begin{itemize}
    \item 1 if r $\leq$ 1
    \item 5 if r $\geq$ 5
    \item r otherwise
\end{itemize}

The only differences between the two experiments consist in how complex categorical features are treated. In the baseline model all features (both for items and for users) are treated as they are in the original data set. While in the stereotype model, complex categorical features are treated via the rating independent stereotypes previously generated in Section~\ref{sec:algorithm}.

As a measure of accuracy of the results, we report the model performance using the Root Mean Squared Error (RMSE) along with the Mean Absolute Error (MAE). In real word recommendation problems, where we usually deal with huge databases, it is crucial to adopt algorithm that is capable of scaling up. In fact, this is one of the advantages of stereotype model. We are reporting the execution speed in seconds as measured on a Intel Core i7 -7700K CPU @ 4.2 GHz with 64.0 GB RAM.


The results reported in Table~\ref {table:Summary} for both experiments, are the average of a 6-fold cross validation. The result highlight the double benefit of a stereotype-based approach: consistence improvement in prediction accuracy in cold start, and an improved computational time, due to the inherent feature space reduction and grouping that stereotype bring to the problem. For example, using NN Regression and stereotypes lead to an accuracy that is higher than XGBoost using standard features (i.e. baseline) with a time saving of over 35 seconds in new user case. This implies that the improvements in using stereotypes are higher than the improvements in increasing the model complexity from a simple linear regression to XGBoost. Therefore, providing more grounded evidence for the use of stereotypes in cold start phases.
  \vspace*{-1.5mm}






%% file: conclusion.tex
\section{Conclusion and Future Work}
\label{sec:conclusion}

In this work, hierarchical clustering of the correlation matrix of complex categorical features is conducted, leading to the formulation of an algorithm for the automatic identification of stereotypes. This can be viewed as a form of feature engineering, where patterns discovered in the correlation matrix are used to drastically simplify a complex categorical feature. The rating independent clusters leads to subjective groups that seem to better partition into classes - stereotypes - than those suggested by the application of k-modes. The lack of a frequency overweight of the labels allows to stereotype all categories, also those that have only a small amount of samples in the data set.

The stereotypes obtained with the proposed algorithm are then used in recommendation. We have shown how, for the experimental MovieLens and Imdb data sets, the stereotype base model outperform a standard feature based recommendation approach for the new user/new item problems. 

Stereotyping numerical features to study the effect of both categorical and numerical stereotypes on the recommendation is the obvious next step. More emphasis should be toward extra reduction in the dimension of the recommendation model, and to improve the quality of recommendations in the new user and new item scenarios (cold start problem). Additional future work will include measuring other metrics including computational efficiency and diversity of recommendation.